\documentclass[11pt,fleqn]{article}
\usepackage{amsmath,graphicx}

\usepackage{amsfonts,amssymb,cite}
\usepackage{epsf,graphicx}

\def\noi{\noindent}

\newcommand{\Title}[1]{\noi {{\Large\bf #1}}\\[1ex]}

\def\Aunames#1{\noi{\bf #1}}
\def\auth#1{${}^{#1}$}
\def\Addresses#1{\medskip\noi \protect
	\begin{description}\itemsep -3pt {\it #1} \end{description}}
\def\addr#1#2{\item[${}^{#1}$]{\it #2}}

\def\nqq{\hspace*{-2em}}

\def\Jl#1#2{#1 {\bf #2},\ }

\def\ApJ#1 {\Jl{Astroph. J.}{#1}}
\def\CQG#1 {\Jl{Class. Quantum Grav.}{#1}}
\def\DAN#1 {\Jl{Dokl. AN SSSR}{#1}}
\def\GC#1 {\Jl{Grav. Cosmol.}{#1}}
\def\GRG#1 {\Jl{Gen. Rel. Grav.}{#1}}
\def\JETF#1 {\Jl{Zh. Eksp. Teor. Fiz.}{#1}}
\def\JETP#1 {\Jl{Sov. Phys. JETP}{#1}}
\def\JHEP#1 {\Jl{JHEP}{#1}}
\def\JMP#1 {\Jl{J. Math. Phys.}{#1}}
\def\NPB#1 {\Jl{Nucl. Phys. B}{#1}}
\def\NP#1 {\Jl{Nucl. Phys.}{#1}}
\def\PLA#1 {\Jl{Phys. Lett. A}{#1}}
\def\PLB#1 {\Jl{Phys. Lett. B}{#1}}
\def\PRD#1 {\Jl{Phys. Rev. D}{#1}}
\def\PRL#1 {\Jl{Phys. Rev. Lett.}{#1}}

\def\lal{&&\nqq {}}

\def\beq{\begin{equation}}
\def\eeq{\end{equation}}
\def\bear{\begin{eqnarray}}
\def\bearr{\begin{eqnarray} \lal}
\def\ear{\end{eqnarray}}
\def\earn{\nonumber \end{eqnarray}}

\begin{document}

\Title{NEW SYMMETRIES, CONSERVED QUANTITIES \\ AND GAUGE NATURE OF A FREE DIRAC FIELD
}

\Aunames{Vladimir V. Kassandrov\auth{a,1} and Nina V. Markova\auth{b,2}}

\Addresses{
\addr a {Institute of Gravitation and Cosmology, Peoples' Friendship
	University of Russia, Moscow, Russia}
\addr b {Department of Applied Mathematics, Peoples' Friendship
	University of Russia, Moscow, Russia}
	}

\abstract{We present and amplify some of our previous statements on  non-canonical interrelations  between the solutions to free Dirac equation (DE) and Klein-Gordon equation (KGE). We demonstrate that all the solutions to the DE (possessing point- or string-like singularities) can be obtained via differentiation of a corresponding pair of the KGE solutions for a doublet of scalar fields. On this way we obtain a  ``spinor analogue'' of the mesonic Yukawa potential and previously unknown chains of solutions to DE and KGE, as well as an exceptional solution to the KGE and DE with a finite value of the field charge (``localized'' de Broglie wave). The pair of scalar ``potentials'' is defined up to a gauge transformation under which corresponding solution of the DE remains invariant. Under transformations of Lorentz group, canonical spinor transformations form only a subclass of a more general class of transformations of the solutions to DE upon which the generating scalar potentials undergo transformations of internal symmetry intermixing their components. Under continious turn by one complete revolution the transforming solutions, as a rule, return back to their initial values (``spinor two-valuedness'' is absent). With arbitrary solution of the DE one can associate, apart of the standard one, a non-canonical set of conserved quantities, positive definite ``energy'' density among them, and with any KGE solution -- positive definite ``probability density'', etc. Finally, we discuss a generalization of the proposed procedure to the case when the external electromagnetic field is present}

{\bf Keywods:} Dirac and Klein-Gordon equations, restricted gauge invariance, spinor two-valuedness, singular solutions, conservation laws, Weyl and wave equations

\newcommand{\be}{\begin{eqnarray}\label}
\newcommand{\ee}{\end{eqnarray}}
\newcommand{\prt}{\partial}
\newcommand{\p}{\prime}
\newcommand{\bib}{\bibitem}
\newcommand{\nabl}{{\bf \nabla}}     
\newcommand{\nd}{{\noindent}}     
\newcommand{\un}{\vskip1cm}      
\newcommand{\hf}{\vskip0.5cm}      
\newcommand{\mb}{\mathbb}      
\newcommand{\wh}{\widehat}
\newcommand{\highlight}{\textcolor{blue}}

\section{On well and not well known: relativistic field equations and equivalence of their  solutions}
Close relationship established in the framework of relativistic field theory between the physical Minkowski space-time geometry $\bf M$ and two observed types of elementary particles, {\it bosons} and {\it fermions}, is, perhaps, the most remarkable achievement in theoretical physics. The connection manifests itself in the (discovered by E. Cartan~\cite[ch. VIII]{Cartan})  existence of two and only two types of irreducible representations of the Lorentz group, symmetry group of $\bf M$, -- {\it tensorial} and {\it spinorial}~ones. According to the modern paradigm, each type of particles of {\it integer} and {\it half-integer} spins (bosons and fermions, respectively) is described by some linear field equation form-invariant under transformations of Lorentz group. In so doing, to ensure the form-invariance, the fields themselves (``wave functions'') should transform through some irreducible representation of Lorentz group (bosons through a tensorial while fermions through a spinorial one, respectively). 

Unconditionally, this rule is fulfilled for particles of low spins. Specifically, spinless bosons ($\pi$-mesons) are described by scalar fields subject to the Klein-Gordon equation whereas particles of unit spin -- to the Proca equations for massive vector field or wave equations for 4-potentials of the massless electromagnetic field. Fundamental fermions (electron, neutrino, proton, neutron, etc.) of spin $1/2$ are described by the Dirac equation whose 4-component field transforms as a bispinor (a pair of 2-spinors)~\footnote{Massless fermions should be described by the Weyl equation. However, after the concept of {\it massive} neutrino had been accepted, corresponding position turned out  to be vacant, and such a situation looks rather strange from a general viewpoint}. 

Dirac equation corresponds to the principal constituents of matter. Introduction of interaction between fermions as matter fields is then realized via the fields of integer spin, carriers of interaction, on the base of the requirement of local gauge invariance of the full Lagrangian. Under gauge transformations the local phase of a wave function and the potentials of a gauge field change concordantly (the latter in a gradient-wise way) while the field strengths remain invariant. However, not all the integer spin fields, the interaction carriers, can be considered as gauge fields; this is, in particular, the case of Klein-Gordon ``mesonic'' field.  In the same manner, the Dirac field is not a gauge one by itself. However, {\it this evidently looking statement will be disputed and refuted further on}. 

As for the full set of {\it solutions} to the equations for free relativistic fields, apart of the most often encountered {\it everywhere regular} solutions of the plane wave type, there exists a wide class of their solutions {\it singular} on a zero measure set, with isolated pointlike or string-like singularities. Besides spherical waves, well known examples of solutions with a pointlike singularity are the Coulomb solution to Maxwell equations or the mesonic Yukawa potential subject to the Klein-Gordon equation. Not always one can put in correspondence with such a singular solution some $\delta$-like {\it source}: for example, this is impossible for the ``flat limit'' electromagnetic field of the Kerr-Newman solution in GTR (the so-called Appel solution~\cite{Witt}) with a ring-like singularity, because of the twofold structure of the solution. 

It is also known that, in the relativistic QM framework, wave functions of the bound $s-$ and $p$- states of the hydrogen atom have a weak singularity in the origin. It turns out that solutions with not only point- or string- but even membrane-type singularities do exist for free Maxwell, Weyl, Yang-Mills fields~\cite{IJGMMP, GR00} and, as we shall show later, for massive Klein-Gordon and Dirac fields as well. It is this, most general class of solutions to relativistic field equations that we shall deal with below. Some of such singular solutions are well known and possess generally accepted physical interpretation; others seem to be novel. 

In the case when {\it all the solutions} to a fundamental field equation can be obtained from solutions of another equaion, and {\it vice versa}, one should not consider such equations as independent and describing {\it different} types of particles. On the contrary, these equations should, perhaps, be treated as mathematically equivalent and {\it corresponding to one and the same physical system in different representations}. Evident example of such a situation is  the link between Maxwell equations for field strengths and wave equations for electromagnetic potentials. Indeed, one can (at least locally) juxtapose to any solution of Maxwell equations  a class of (gauge equivalent) potentials subject to wave equations, and conversely. That is why we do not relate, of course, these equations to different physical entities but consider them both as describing one and the same electromagnetic field in its different representations. 

It turns out that similar {\it equivalence} relationship takes place between the solutions to massless ( Maxwell, Weyl and d'Alambert)~\cite{Sing,Equiv} and massive (Dirac and Klein-Gordon)~\cite{Equiv,Zommer} equations describing as though {\it different} types of particles. Evidently, the arising situation hardly corresponds to the generally accepted viewpoint. Indeed, in the case of, say, massive fields it is usually postulated that the Dirac equation (DE) and the equation of Klein-Gordon (KGE) are responsible for description of particles of different spins, possess different sets of conserved quantities and transform through different representations of the Lorentz group. As for their solutions, everybody knows that each component of the Dirac field identically satisfies the KGE, but not vice versa ! In this sense the DE is usually considered more {\it rigid} and informative than the KGE ~\cite[p. 56]{Bogol}. 

Meanwhile, correspondence of solutions to the DE and KGE can be not algebraic but {\it differentiative} in nature just as it takes place for Maxwell equations for strengths in compare with wave equations for potentials. Indeed, as it has been proved in ~\cite{Equiv} (see also~\cite{Zommer}), {\it any solution to the DE can be obtained by differentiation of a corresponding set of four solutions to the KGE} (defined up to a specific ``gauge'' freedom, see below).

In other words, one can regard a quadruple of the Klein-Gordon fields as a sort of ``potentials'' for the Dirac bispinor field~\cite{Equiv}. From this point of view the {\it free Dirac field is itself a gauge field},  and the DE and KGE should be considered as mathematically equivalent and describing one and the same type of particles. Corresponding construction proposed earlier in ~\cite{Equiv} is presented in Section 2. In Section 3, by making use of the 2+2 representation of the DE,  this construction is refined and essentially reinforced. It is shown, in particular, that {only some two (not four !) solutions to the KGE are enough to obtain through differentiation an arbitrary solution to the DE}.

As far as the components of the arising solutions to the DE, in the turn, also satisfy the KGE, it becomes possible to generate a whole {\it chain} of the DE-KGE solutions. Such a possibility is demonstrated in section 4 on a number of examples for which, as a starting point, stationary (or static) spherically (axially) symmetric solution to the KGE, with point- or string like singularities is taken. In particular, a ``spinorial analogue'' of the   mesonic Yukawa potential is obtained. At the end of Section 4, a simple solution to the KGE with integrable singularity and finite value of the field charge , as well as its ``spinorial  analogue'' is presented. Solution with such an exceptional property seems to be discovered for the first time among those of various linear equations for fundamental fields. After a corresponding Lorentz boost, we are brought up to its natural interpretation as a {\it ``localized'' de Broglie wave}.     

On the other hand, the discovered possibility to obtain {\it general solution to the DE} from the solutions for {\it scalar} fields poses, in the framework of the considered approach, the question about the origin of the {\it spinor} law of transformation of the Dirac field. Solution of this problem had been proposed in ~\cite{Equiv} and is based on the use of the {\it internal symmetry} of the KGE system with respect to the transformations of the group $SL(4,\mathbb C)$, {\it intermixing} the components of the quadruple of the Klein-Gordon ``potentials''. After an appropriate ``tuning'' of such transformations to the transformations of Lorentz group the canonical spinor law of transformations of the Dirac field can be completely restored.  Generally, however, the arising law of transformations of the DIrac field components corresponds to a {\it nonlinear} representation of the Lorentz group and, on the other hand, does not result in a spinorial two-valuedness. In particular, {\it after a rotation by the complete angle the Dirac field under transform does not return, as a rule,  back to its initial values}. Properties of extended symmetry transformations of the DE-KGE solutions are examined in Section 5 and illustrated on the above indicated examples of solutions from Section 4.  

In Section 6 we examine the problem of ``twofoldness'' of the set of conserved quantities which can be juxtapose to any solution to the DE or the KGE owing to their mutual correspondence. To the first turn it is remarkable that there exist two different ``energies'' of the Dirac field one of which, according to the properties of the associated  scalar fields, is positive definite! Conversely, to any solution of the KGE a positive definite ``probability density'' can be ascribed! 

In Section 7 we briefly describe the situation arising in the attempts of generalization of the presented construction to the case when an external electromagnetic field is present. Massless case is briefly reviewed in Section 8. In conclusion (Section 9) the most important results of the paper are summarized. The problem of physical interpretation of the established equivalence properties for distinct relativistic field equations is touched upon as well as possible consequences of this equivalence for the QFT.  Finally, in the appendix, we present the action for the KGE potentials of the Dirac field. It is Lorentz and gauge invariant and leads to the equations of the third order in derivatives which are satisfied by the solutions of both DE and KGE. 

To simplify the perception, in the main part of the paper we do not apply the 2-spinor formalism but use instead an equivalent 2+2 matrix form of representation. For  metric on $\bf M$ the form  $\eta_{\mu\nu}=diag\{+1,-1,-1,-1\}$ is chosen so that, say, the d' Alembert operator has the form  $\square:=-\partial_\mu\partial^\mu=\Delta - \partial^2/\partial t ^2$. As usual, system of units where $c=1,\hbar=1$ is used throughout the paper.   

\section{Klein-Gordon ``potentials'' and gauge nature of a free Dirac field}
Consider the Klein-Gordon equation (KGE)
\be{column}
(\square - m^2)\Phi = 0 ,
\ee
for four free complex scalar fields  $\Phi=\{\phi_a\},~~a=1,2,3,4$. 
The Klein-Gordon operator can be factorized 
\be{factor}
(\square - m^2) = DD^* = D^*D,
\ee
into the product of two commuting Dirac operators $D,D^*$ of the first order: 
\be{oper}
D:= i\gamma^\mu \prt_\mu - m,~~~D^*:=i\gamma^\mu \prt_\mu + m,
\ee
where $\gamma^\mu$  are the $4\times 4$  Dirac matrices 
\be{commut}
\gamma^\mu \gamma^\nu + \gamma^\nu \gamma^\mu = 2\eta^{\mu\nu}, 
\ee
which we shall take in the standard 2+2 representation making use of the Pauli matrices  (see below, Section 3).

Through the derivatives of $\Phi$ let us then define another 4-component complex field  $\Psi$,  
\be{poten}
\Psi:=D^*\Phi, 
\ee
which represents, on account of (\ref{column}) and (\ref{factor}), a solution to the Dirac equation (DE),  
\be{dirac}
D\Psi = DD^*\Phi = 0.
\ee

Conversely, let an arbitrary solution $\Psi$ to the DE, $D\Psi=0$, is given\footnote{Then, as is well known, each component $\psi_a$  satisfies identically the KGE,  since $0=D^*(D\Psi) = (\square-m^2)\Psi = 0$}. In this case, the system of four inhomogeneous first order equations (\ref{poten}) can be always (locally) resolved with respect to four unknowns $\phi_a$. As a partial solution of (\ref{poten}), on account of the identity $D^*=D+2m$, one can take the Dirac field itself,
\be{soleq}
\Phi=\frac{1}{2m}\Psi,
\ee
 while more general solution to (\ref{poten}) is represented (on account of $D^*\gamma^5 = -\gamma^5 D$) by the ``projective'' ansatz
\be{solpart}
\Phi = \frac{1}{2m}(1\pm \gamma^5) \Psi, ~~\gamma^5:=i\gamma^0\gamma^1\gamma^2\gamma^3,
\ee
that is, by ``right'' or ``left'' 2-spinors corresponding to the bispinor $\Psi$. Thus, {\it any solution of the DE can be in fact obtained from only {\it a pair} of the Klein-Gordon fields} (for more details, see Section 3). 

Of course, the obtained functions $\phi_a$ are subject to the KGE,
\be{ident2}
0=D\Psi = DD^*\Phi=(\square-m^2)\Phi \equiv 0, 
\ee
yet defined non-uniquely, up to the general solution of an {\it homogeneous} equation of the type (\ref{poten}). Specifically, any initially fixed solution $\Psi$ of the DE  (``strengths'' of the Dirac field) remains invariant under the following {\it gauge} transformations of the corresponding ``Klein-Gordon potentials'' from (\ref{poten}): 
\be{gauge}  
\Phi \mapsto \Phi + \Upsilon, 
\ee
with $\Upsilon$ being some {\it arbitrary} solution of the (conjugate) DE, 
\be{dirac2}
D^*\Upsilon = 0. 
\ee
In so doing, since for each $\Upsilon$ some Klein-Gordon potentials $\Xi$ do exist, that is $\Upsilon = D\Xi$, the gauge transformation  ({\ref{gauge}) can be represented in a familiar gradient-wise form
\be{grad}
\Phi \mapsto \Phi + D\Xi. 
\ee
Thus, for any DE solution potentials subject to the KGE are locally defined up to the gauge transformations (\ref{grad}). Through their differentiations {\it the complete set of solutions to the free DE can be obtained}~\cite{Equiv}. As to the DE itself, {\it free Dirac  field should be regarded as a gauge field}, in full analogy with free Maxwell equations for the strengths of electromagnetic field.

However, the set of four gauge functions $\Xi=\{\xi_a\}$, in distinction with the gauge symmetry in electrodynamics, is not arbitrary but subject to the KGE, $(\square-m^2)\Xi =0$. This resembles the ``residual'' gauge invariance
\be{maxgauge}
A_\mu \mapsto A_\mu -\prt_\mu \alpha,~~~~\square \alpha =0
\ee
of Maxwell equations $\prt^\nu F_{\mu\nu}=0,~F_{\mu\nu}=\prt_\mu A_\nu-\prt_\nu A_\mu$, {\it supplemented by the Lorentz gauge equation for the potentials  $\prt_\mu A^\mu =0$}. 

It is also worthy to note that similar ``weak'' gauge invariance (for which the gauge function can depend on coordinates {\it implicitly}, only through the components of the field function under transform) takes place for the class of solutions to relativistic field equations generated by twistor functions (for detail, see~\cite{IJGMMP}).

\section{Any solution to free Dirac equation 
from a doublet of the Klein-Gordon scalar fields}
Many of the interrelations between the Dirac and Klein-Gordon fields, afore presented among them, look more transparent in the {\it chiral} representation of the DE. Specifically, let us define the $2\times 2$ matrix-valued {\it Weyl operators} (principal and conjugate, respectively): 
\be{weyl}
W:=(\prt_t - \vec \sigma \nabla), ~~ \tilde W:=(\prt_t + \vec \sigma \nabla)
\ee
where $\vec \sigma =\{\sigma_a\},~a=1,2,3$  are the Pauli matrices. Then the DE 
\be{Dirac2}
D\Psi: = (i\gamma^\mu \prt_\mu -m)\Psi =0.
\ee
aquires the following ``splitted'' form (see, e.g.,~\cite[ch.II, section 9]{Akhiezer}):
\be{Dirac2+2}
W a = -i m b, ~~ \tilde W b = - i m a, 
\ee
where the 2-component ``right'' $\Psi_R:=a$ and ``left'' $\Psi_L:=b$ spinors 
are defined as the half-sum / half-difference of the initial Dirac 2-spinors $\psi^T =\{\kappa,\chi\}$, 
\be{2-spinors}
a=(\kappa+\chi)/2,~~~b=(\kappa-\chi)/2 ,  
\ee 
Note that Weyl operators $W,\tilde W$ are Hermitian and factorize the d'Alambert wave operator 
\be{dalamber}
W\tilde W = \tilde W W=-\square =\prt^2/\prt t^2 - \Delta. 
\ee 
The above described procedure to seek the Klein-Gordon potentials based on resolution of  equations (\ref{poten}), in the chiral representation reduces to resolution of the following system of equations 
\be{poten2+2}
a = \tilde W \beta - i m \alpha, ~~~ b = W \alpha - i m \beta
\ee
with respect to a pair of unknown 2-component functions $\{\alpha,\beta\}$ for any given 2-spinors $\{a,b\}$ subject to the DE (\ref{Dirac2+2}). It is easy to check that whether the solution to (\ref{poten2+2}) does exist, then 1) it is non-unique and 2) the potentials $\{\alpha,\beta\}$, on account of (\ref{Dirac2+2}) and (\ref{dalamber}), should satisfy the KGE, $(\square-m^2)\alpha=0$,~~$(\square-m^2)\beta =0$. Specifically, the gauge transformation of potentials which leave invariant both Dirac 2-spinors has the form 
\be{gauge2+2}
\begin{array}{cc}
\alpha \mapsto \alpha -m^2 \pi - i m \tilde W \rho, \\
\beta \mapsto \beta + m^2 \rho + i m W \pi. 
\end{array}
\ee
In the above equations $\pi,\rho$ are two arbitrary and independent pair of functions each component of which satisfies the KGE.

Now the problem of {\it existence} of potentials subject to  (\ref{poten2+2}) can be explicitly resolved. Indeed, let us nullify\footnote{This evidenly corresponds to solutions (\ref{solpart}) of (\ref{poten}) for potentials $\Phi$  in the initial 4D representation, see Section 2} one of the 2-component potentials $\{\alpha,\beta\}$  setting, say, $\beta=0$.   Then the system (\ref{poten2+2}) reduces to {\it identification} of the other 2-component $\alpha$ with the first of the given 2-spinors 
\be{first2sp}
\alpha = \frac{i}{m} a, 
\ee
while the other 2-spinor is then expressed through the derivatives of the first one:
\be{second2sp}
b=\frac{i}{m} W a. 
\ee   

In other words, for any solution to the DE the first of equations (\ref{poten2+2}) is simply a definition ({\ref{second2sp}) of the second 2-spinor $(b)$ through the first one $(a)$, after which the second equation is identically valid since the KGE holds for both components of the principal 2-spinor $(a)$. 

Thus, {\it any solution to  DE is represented by some two functions, say, $(a)$ subject to the KGE} which define themselves one of the Dirac 2-spinor whereas the second 2-spinor $(b)$ is explicitly expressed through the derivatives of the first one $(a)$! In the next Section we shall exhibit simple examples of such a procedure and obtain a number of singular solutions to the DE.  
 
   \section{Chains of singular solutions to the Dirac and Klein-Gordon equations}
 In the above described method of generation of the DE solutions from some pair of the KGE solutions, components of the arising Dirac fields, as it usually is, satisfy the KGE themselves and, therefore, can serve as ``potentials'' to obtain new soluions to the DE, and so on. A chain of the DE-KGE solutions arising in such a way proves to be infinite or terminates in the case when new solutions turn to be functionally dependent on the old ones. 
 
On account of {\it linearity} of the considered equations it is sufficient to restrict oneself by the case when only one of the initial KGE solutions is nonzero and take, say, the initial 2-spinor in the form $a^T = (0, F)$ or $a^T=(G,0)$, where functions $F,G$  represent some solutions to the KGE. Then general case is represented by the 2-spinor $a^T=(G,F)$, which generates a superposition of the DE solutions related to ``partial'' 2-spinors.  

As generating potentials we below consider stationary (in particular, static) solutions of the KGE possessing spherical or axial symmetry. Specifically, for stationary solutions 
\be{station}
F=f(\vec r) e^{-i\omega t}
\ee
the KGE reduces to the form 
\be{KGstation}
\Delta f +(\omega^2 - m^2)f=0, 
\ee
and, in the most interesting case $\vert \omega \vert =m$, -- to the Laplace equation $\Delta f =0$. Selecting as the solutions of the latter the {\it Coulomb field} and the {\it stereographic projection}, one comes to the following stationary solutions to the KGE:
\be{CoulombKG}
F=\frac{1}{r} e^{\pm i m t},
\ee
and
\be{StereoKG}
F=\frac{x+iy}{r+z}e^{\pm i m t},
\ee 
 respectively ($x,y,z$ being the Cartesian coordinates, and $r=\sqrt{x^2+y^2+z^2}$).   
         
First of the above presented solutions has a pointlike singularity, while the second -- singularity of a string type (on the negative symmetry semi-axis). As it had been already
mentioned in the introduction, such solutions are valid only in the external space {\it outside  singular domains} and, generally, it does not look justified to define corresponding singular ``sources''. For example, in the {\it whole} space solution (\ref{CoulombKG}) will satisfy the {\it inhomogeneous} KGE with ``oscillating'' pointlike source of the form $\delta(\vec r)e^{\pm i m t}$ looking physically senseless. The same can be said on the introduction of a source localized on the singular string of the soluiton (\ref{StereoKG}). Remarkably, Dirac himself had no intention to introduce such a source on the singular string of the Schrodinger's wave function of the electron in the field of a magnetic monopole~\cite[Sect. 4]{DiracString}. 
      
Consider now the other limiting case of static solutions to (\ref{KGstation}) with $\omega=0$. Let us present here the following two static solutions of the KGE:
\be{Yukawa}
F=-\frac{g^2}{r}e^{-mr}
\ee 
and
\be{SterCoul}
F=\frac{x+iy}{r(r+z)} e^{-mr}
\ee
first of which is known as the {\it Yukawa potential} in the old theory of mesonic forces (with $g^2$ being the interaction constant). 

Each of the two above presented solutions to the KGE can be completed up to a corresponding solution of the DE by taking, say, the first 2-spinor $(a)$ in the form $a^T=(0,F)$ and defining the second 2-spinor $(b)$ in accordance with (\ref{second2sp}). In such a way, making use of the generating solution to the KGE (\ref{Yukawa}), one obtains, in particular, the ``spinorial analogue'' of the Yukawa potential, that is, the solution to the DE of the form
\be{YukawaSpinor}
 a=-\frac{g^2}{r}e^{-mr} \left(\begin{array}l
0 \\ 
1
  \end{array}
  \right), 
 ~b=-\frac{ig^2}{mr^3}(1+mr)e^{-mr}\left(\begin{array}l
x-iy\\ 
-z
  \end{array}
  \right).  
 \ee
  Selecting then one of the components of the obtained 2-spinor $(b)$ as an initial generating function $F$ instead of (\ref{Yukawa}) and using again the formula (\ref{second2sp}), one obtains the next solution to the DE in arising infinite chain of solutions:
\be{yukava_ab1}
a=\left(\begin{array}l
~~~~~~~~0 \\ 
\frac{x-iy}{r^3}(1+mr)e^{-mr}
  \end{array}
  \right), 
 ~b=-\frac{i}{mr^5}(3+3mr+m^2r^2)e^{-mr}\left(\begin{array}l
~~(x-iy)^2\\ 
-(x-iy)z
  \end{array}
  \right).  
 \ee
Note that the components of Dirac fields entering the solutions, from the viewpoint of their {\it  angular dependence}, are various combinations of {\it spherical spinors} (see, e.g.,~\cite[ch.2, sect. 11]{Akhiezer}). On the contrary, for the DE solutions with stringlike singularities generating by (\ref{StereoKG}) or (\ref{SterCoul}),  angular dependence  
seems to be nontrivial and needs a further study.  

To conclude, let us present a simple yet {\it exceptional in its properties} one-parameteric family of stationary spherically symmetric solutions to the KGE which can be easily obtained by separation of variables in (\ref{KGstation}). It has the following form:
\be{BroglieKG}
F=\frac{e^{-mkr}}{r} e^{-im\omega t}, ~~~k^2+\omega^2 =1, 
\ee  
or, using evident parameterization $k=cos{\psi},\omega=sin{\psi}$, 
\be{dBroglieKG}
F=\frac{e^{-mr cos{\psi}}}{r} e^{- imt sin{\psi} }. 
\ee     
As the Yukawa solution, function (\ref{BroglieKG}) is square integrable. Remarkably, however, the associated  charge-current density 4-vector 
\be{currntKG}
J_\mu =:\frac{i}{2}(F^*\prt_\mu F -\prt_\mu F^* F),~~\prt_\mu J^\mu =0, 
 \ee 
on the soluiton (\ref{BroglieKG}), {\it defines a finite conserved quantity, the ``field charge''} 
\be{charge}
Q=\frac{1}{4\pi}\int J^0 dV = m\omega \int_0^\infty \vert F \vert^2 r^2 dr = \frac{\omega}{2k} = \frac{1}{2}\tan{\psi}.   
\ee
To our knowledge, {\it the above obtained solution with a finite integral of motion  is the only known one among the solutions to all linear equations for free relativistic fields}.  For $\omega=0~(\psi=0)$ it reduces to the static Yukawa solution with zero field charge while for $k=0~ (\psi=\pm \pi/2)$ -- to the ``long-range'' solution 
(\ref{CoulombKG}) for which the integral $Q$ diverges at spacial infinity. 

Solution (\ref{BroglieKG}) can be also ``completed'' to a DE solution, again by making use of the formula (\ref{second2sp}). Corresponding ``spinorial analogue'' of the solution  (\ref{BroglieKG}) is of the form:
\be{spinorBroglie}
a=\frac{e^{-mkr-im\omega t}}{r}\left(
\begin{array}l
0 \\
1
\end{array}
\right) , 
~b=-i\frac{e^{-mkr-im\omega t}}{mr^3}\left(
\begin{array}l
-(x-iy)(1+mkr) \\
z(1+mkr)+im\omega r^2
\end{array} \right), 
\ee
where, as before, $k^2+\omega^2 =1$.

The canonical ``Dirac'' field charge with positive definite density $\sim (a^2+b^2)$ will be infinite (divergence in the singular point). Nonetheless, since the nonzero component $F$ of the spinor $a$ satisfies the KGE, the charge $Q$ it associates can be treated as the integral of motion with respect to the Dirac field as well (for more detail see Section 6). As to other solutions in the chain generated by (\ref{BroglieKG}), they do not possess a finite-valued  charge at all.  

Evidently, solution obtained from (\ref{spinorBroglie}) by a ``boost'' can be interpreted as a {\it ``localized'' de Broglie wave} with a finite-valued field charge.

\section{Spinors from scalars: non-canonical symmetries of a free Dirac field}
In the above presentation we considered, as is generally accepted, both 2-component functions ($a$ and $b$) as 2-spinors. On the other hand, the generating KGE solutions which form the ansatz $a$, according to their internal properties, under transformations of Lorentz group should be treated as scalars. Thus, one encounters the problem: in which way the scalar nature of the initial Klein-Gordon fields can be put in correspondence with the spinor law of transformation of the Dirac field? 

To most part, this problem had been already solved in ~\cite{Equiv}, and below we elucidate the solution on the base of the chiral representation (\ref{Dirac2+2}) of the DE and the above considered examples of solutions. 

Under (proper) Lorentz transformations of coordinates
\be{lorentz}
X \mapsto \bar X= S X S^+,  
\ee 
where 
\be{coord}
X=X^+ =t+\vec \sigma \vec r
\ee
is the Hermitian matrix of coordinates on $\bf M$, the Weyl operator $W$ (and its conjugate  
 $\tilde W$) are transformed as 
\be{weyltransform}
W \mapsto \tilde S^+ W \tilde S, ~~\tilde W \mapsto  S \tilde W S^+,   
 \ee    
where $S\in SL(2,\mathbb C)$  is an arbitrary matrix from Lorentz ``spinor group''  with  ``half-angles''  of (pseudo) rotations as 6 parameters representing (up to a sign) an arbitrary Lorentz transformation from the $SO(3,1)$ group; here $\tilde S, S^+$  are matrices inverse and Hermitian conjugate to $S$, respectively. 

According to the canonical spinor law, corresponding to (\ref{lorentz}), transformations of the quantities $a,b$  have the form 
 \be{sptransform} 
 a(X) \mapsto  \bar a (\bar X) =S a(X), ~~~b(X) \mapsto \bar b (\bar X) =\tilde S^+ b( X),
 \ee
so that the DE system (\ref{Dirac2+2}) remains form-invariant, and one obtains a solution to the DE $\{\bar a,\bar b\}$ corresponding to the initial one $\{a,b\}$ transformed to a new reference frame. 
 
On the other hand, treating the initial components of $a$ subject to the KGE as scalars, one should transform only their arguments, 
 \be{transformarg}
 a (X) \mapsto \bar a (\bar X) =a(X), 
\ee    
after which for the components of $b$, according to (\ref{second2sp}), one obtains the expression
 \be{transform_nl}
 b(X) \mapsto \bar b (\bar X)= \frac{i}{m} (\tilde S^+ W \tilde S) a(X),
 \ee
so that the pair (\ref{transformarg}),(\ref{transform_nl}) represents a new solution to the DE $\{\bar a,\bar b\}$ {\it generally distinct} from that canonically transformed one (\ref{sptransform}). Note that the new components of $\bar b$, according to  (\ref{transform_nl}), cannot be expressed {\it algebraically} through the initial functions $b$ so that the considered transformations define a {\it nonlinear representation} of the Lorentz group. 

In fact, actual possibility of the two different types of symmetry transformations of the DE solutions is related to the existence of a supplementary {\it internal} symmetry of the KGE for the doublet of scalar fields $a(X)$ w.r.t. transformations from the group $SL(2,\mathbb C)_{(INT)}$ which represents an independent {\it copy} of the spinor Lorentz group $SL(2,\mathbb C)$. Such transformations do not change the coordinates themselves but linearily intermix the components of the initial scalar doublet,   
\be{transf_int}
a(X) \mapsto \bar a = M a(X), ~~~M\in SL(2,\mathbb C)_{(INT)}. 
\ee
Combining now these transformations with those from the spinor Lorentz group, we obtain {\it the most general law of transformations for the DE solutions} in the following form:
\be{transform_gen}
\bar a (\bar X) = M a (X), ~~~\bar b (\bar X)= \frac{i}{m} (\tilde S^+ W \tilde S M) a(X).
\ee

In general, parameters of the matrix $M$ are entirely independent from those of Lorentz transformations. On the other hand, whether one identifies $M\equiv S$, the canonical spinor law of transformations (\ref{sptransform}) will be restored:
 \be{restor_transform}
 \bar a = Sa(X), ~~~\bar b = \frac{i}{m} (\tilde S^+ W) a(X) \equiv \tilde S^+ b(X).
 \ee
Note, however, that generally, if only the parameters of a matrix $S$ corresponding to ``half-angles'' of a 3D rotation do not enter the matrix $M$, the starting  solution to DE, being continiously transformed according to (\ref{transform_gen}) {\it returns back to its initial value after one complete revolution}; in other words, the customary {\it spinor two-valuedness is generally absent}. From a physical viewpoint, all the KGE-DE solutions obtained from an initial one by means of a combination of Lorentz transformations with transformations of the internal group, should be, perhaps, regarded as {\it equivalent}.
\vskip2mm

Let us now illustrate the above described scheme on the examples of solutions to DE written out  in Section 4. Below we do not deal with arbitrary ``intermixings'' of the components (\ref{transform_gen}) but restrict our consideration by two limiting cases comparing the {\it canonical} spinor transformation (\ref{sptransform})  with {\it alternative} (scalar with respect to the components of $a(X)$}) transformations (\ref{transformarg}),(\ref{transform_nl}).  

It is easy to check that, say, spherical symmetric (by norm) ``spinor Yukawa solution''  (\ref{YukawaSpinor}) as well as the exceptional solution (\ref{spinorBroglie}) or that generated from (\ref{CoulombKG}), under rotation by an angle $\varphi$ round an arbitrary axis, transform in a rather trivial way. Specifically, according to the canonical law of transformations, all spinor components of those solutions acquire common ``spinorial'' phase factor $\exp{(- i \varphi/2})$. Under alternative transforsformation the components of both solutions {\it do not change at all} so that one encounters here examples of the 
{\it entirely $SO(3)$- invariant, ``scalar-like''  (w.r.t. the alternative transformations) solutions to the DE} for which $\bar a(\bar X)=a(X),~\bar b(\bar X) = b(X)$! Similarly, the third spherically symmetric ``by norm'' stationary solution to the DE which can be obtained from the generating KGE solution (\ref{SterCoul}) is also $SO(3)$-invariant.  

As for the axisymmetric solutions to DE which correspond to generating scalar functions (\ref{StereoKG}) or (\ref{SterCoul}), they acquire the common ``spinorial '' factor $\exp{(+i \varphi/2})$ being transformed in the canonical way.  On the other hand, under the alternative transformation both their components become multiplied by  the {\it ``vector-like''} factor $\exp{(+ i \varphi})$ and, after a rotation by  $360^0$, return back to their initial values.  

Consider now one more example of a nontrivial transformation of a solution to DE in the process of which  the initial symmetry is broken. For the ``spinor Yukawa'' solution (\ref{YukawaSpinor}) generated by the spherically symmetric function (\ref{Yukawa}) let us take, as such a transformation, a Lorentz {\it boost} along, say,  $Z$-axis with the velocity parameter $V=\tanh{\theta}$. In this case transformation matrix $S$ has the form 
\be{matlortransf}
S=\left(
\begin{array}l
e^{-\theta/2}~~ 0 \\    
0 ~~~~~e^{\theta/2}  
\end{array}
\right),
\ee 
and for the solution transformed {\it canonically} according to  (\ref{sptransform}) to the new reference frame we obtain (using for simplicity for {\it new} coordinates the same notation as for the initial ones):
\be{lortransf}
\bar a=\frac{e^{\theta/2}}{r_*}e^{-mr_*}\left(\begin{array}l
0 \\ 
1
\end{array}
  \right),  ~~\bar b = \frac{i e^{\theta/2}}{mr_*^3}(1+mr_*)e^{-mr_*}\left(\begin{array}l
~~~~~~~w\\ 
-(1+e^{-2\theta}) z_*
  \end{array}
  \right),  
 \ee
where the following familiar quantities are introduced:
\be{denote}
z_*:=z-Vt,~~~r_*:=\sqrt{x^2+y^2+z_*^2 \cosh^2 \theta}. 
\ee
We see that, apart of the common spinor factor  $e^{\theta/2}$ and canonical transformation of arguments, in the second component of the spinor $b$ there arises a supplementary ``deforming'' factor of the form
\be{deform}
(1+e^{-2\theta})\equiv \frac{1+V}{1-V}.
 \ee
On the other hand, under {\it alternative} transformation of the same spherically symmetric solution, in the single component of $a$ only its argument $r$ should be changed for $r_*$ as in (\ref{denote}).  Computing then, explicitly through $\bar a$ or making use of the law of transformation (\ref{transform_nl}), two novel components of $\bar b$, one results in the Yukawa solution (\ref{YukawaSpinor}) transformed alternatively to the moving reference frame:
\be{altlortransf}
\bar a=\frac{1}{r_*}e^{-mr_*}\left(\begin{array}l
0 \\ 
1
\end{array}
  \right),  ~~\bar b = \frac{i}{mr_*^3}(1+mr_*)e^{-mr_*}\left(\begin{array}l
~~~~~~~~x-iy \\ 
-(1+e^{-2\theta}) z_*
  \end{array}
  \right),  
 \ee
In distinction with the scalar type of transformation of functions $a$, corresponding functions $b$ are again deformed by the arising factor (\ref{deform}), so that the alternatively transformed solution (\ref{altlortransf})  {\it completely reproduces the canonically transformed one} (\ref{lortransf}),  disregarding the absence of the common characteristic ``spinor'' factor $e^{\theta/2}$. Remarkably, under a boost along their symmetry $Z$-axis in a quite analogous way are related the canonically and alternatively transformed axisymmetric solutions to DE generated by functions (\ref{StereoKG}) and (\ref{SterCoul}). 

Let us note in conclusion that, apart of the above considered symmetrical cases, application of the two distinct rules of transformation to an initial DE solution, namely  (\ref{sptransform}) and (\ref{transform_nl}), generally leads to two {\it essentially} different (that is, different not only by the presence/absence of the common spinor factor)  solutions to the DE. One can convince in this fact considering, say, a rotation of an axisymmetrical solution round an axis not coinciding with its initial symmetry axis ($Z$).


\section{Positive definite ``energy'' for Dirac field and ``probability density'' for the Klein-Gordon field} 
In consequence of the mutual correspondence of the DE-KGE solutions, for an arbitrary Dirac field,  apart of the canonical set of integrals of motion, there exists the other one which is defined by corresponding pair of solutions to the KGE, and conversely. This allows, in particular, for the existence of a {\it second}, positive definite ``energy''  density for free Dirac fields, as well as positive definite ``probability'' density for any solution of the KGE.  In a more formal way, the above correspondence can be established in the framework of the Lagrangian approach as follows. 

In the 2+2 representation the Dirac equations (\ref{Dirac2+2}) can be obtained through variation of the Lagrangian
\be{lagrDirac}
L = i \{a^+ (Wa) +b^+ (\tilde W b) - (Wa^+) a - (\tilde W b^+) b\} -2m(a^+ b +b^+ a)
\ee
(in view of  hermiticity of the Weyl operators $W^+ = W,~\tilde W^+ = \tilde W$).

Assuming the fulfilment of DE (\ref{Dirac2+2}) and substituting the derivatives in (\ref{lagrDirac}), one comes to the known property of the Dirac Lagrangian to vanish on the solutions.  On the other hand, exchanging in (\ref{lagrDirac}) the fields $a,b,a^+,b^+$ themselves with corresponding,  on account of (\ref{Dirac2+2}), derivatives one obtains the Lagrangian for {\it two doublets} of fields 
\be{lagrKlein}
L=\frac{2}{m} \{(Wa^+) (\tilde Wb) + (\tilde W b^+) (Wa) -m^2 (a^+ b +b^+ a)\}, 
\ee
variation of which results in the KGE for each component of $\{a,b\}$,  
\be{KGE}
(\square - m^2) a = 0, ~~~(\square - m^2) b = 0 
\ee
and for their hermitian conjugate. Making now use of the standard procedure, for the initial Dirac Lagrangian one defines a canonical set of combinations of field quantities that satisfy on the solutions the {\it continuity equations}. In particular, for the known {\it Dirac charge-current} 4-vector
\be{4-vectorDirac}
j_\mu^{(D)}:= \bar \psi \gamma_\mu \psi,~~~\prt^\mu j_\mu^{(D)} =0, 
\ee
with a positive definite ``charge density'' -- {\it probability density}
 $\rho^{(D)}:=j_0^{(D)} = \psi^+ \psi$, in the 2+2 representation one gets
\be{chargedensD}
\rho^{(D)} =2(a^+a+b^+ b).
\ee      
However, for the same solutions $\{a,b\}$ of the DE-KGE, making use of the Lagrangian (\ref{lagrKlein}), one obtains the expression for a conserved {\it Klein-Gordon current} 4-vector standard for scalar fields:
\be{4-vectorKlein}
j_\mu^{(KG)} =\frac{i}{2} (a^+ \prt_\mu a - \prt_\mu a ^+ a + b^+ \prt_\mu b - \prt_\mu b ^+ b), 
\ee    
which defines a {\it sign-indefinite density of the  ``field charge''} 
\be{chargdensKG}
\rho^{(KG)} = \frac{i}{2} (a^+ \prt_t a - \prt_t a ^+ a + b^+ \prt_t b - \prt_t b ^+ b).
\ee
One can explicitly observe the difference of these expressions, in particular, for the spherically symmetric solutions  to the KGE (\ref{Yukawa}) and the DE (\ref{YukawaSpinor}) related to the Yukawa potential. Indeed, for this solution the density of the field charge ``{\it a l\' a} Klein-Gordon''   (\ref{chargdensKG}) turns to zero while the probability density  (\ref{chargedensD}) is positive and equal to 
\be{staticdens} 
\rho^{(KG)}=\frac{1}{r^2} e^{-2mr}\left(1+\frac{(1+mr)^2}{(mr)^2}\right).
\ee
In general case of {\it stationary} solutions to the KGE-DE with $a,b \sim \exp{(-i \omega t)}$ expressions for two conserved densities are proportional to each other,  
\be{chargdens_statnr}
\rho^{(D)} =2(a^+a+b^+ b),~~\rho^{(KG)} =2\omega(a^+a+b^+ b),  
\ee
but the sign of the second density can be choosed negative, so that the DE can in fact describe the particles with opposite ``charges''. 

Consider now the problem of ``two energies'' for the solutions to the KGE-DE. Let one has a solution to the DE $\{a,b\}$. Then, making use of the canonical form of the energy-momentum tensor of the Dirac field, for its $(00)$-component -- energy density $\epsilon$, in the 2+2 representation one obtains the expression
\be{enrgdensD}
\epsilon^{(D)} = \frac{i}{2} (a^+ \prt_t a - \prt_t a ^+ a + b^+ \prt_t b - \prt_t b ^+ b),
\ee
reproducing that for the Klein-Gordon charge density (\ref{chargdensKG})  and, certainly, sign-indefinite.  However, from corresponding solutions to the KGE and  Lagrangian  (\ref{lagrKlein}) one defines the second ``energy'' density for the same Dirac field: 
\be{enrgdensKG}  
\epsilon^{(KG)}=(\nabla a^+ \nabla a + \prt_t a^+ \prt_t a +m^2 a^+ a) + (\nabla b^+ \nabla b + \prt_t b^+ \prt_t b +m^2 b^+ b), 
\ee
which, of course, is positive definite!
 
Quite analogously, for any KGE-DE solution, apart of the canonical one, it is possible to define {\it another density of the angular momentum} making use of the expression for the second field corresponding to the first one. To be sure, this procedure does not directly relate to the generally accepted opinion on the one-half spin of Dirac particles or zero spin of particles described by the KGE.
         
\section{On correspondence of Dirac and Klein-Gordon equations in an external electromagnetic field}
One can try to generalize the above obtained 
 static or stationary solutions, spherically or axisymmetric, for description of fields produced by a point-like or string-like singularity moving along an {\it arbitrary} world line, in full analogy with the Lienard-Wiehert fields in the massless electromagnetic case. However, even in the case of inertial movement the already obtained solutions allow for a recurrence to the de Broglie's interpretation of the wave-particle duality as the concordant motion of a particle-singularity and a ``pilot wave''~\cite[ch. IX]{Broglie}.  These questions require special consideration. 

As for generalization of the presented construction to the case when the external field is present, electromagnetic or gravitational, one encounters rather obvious problems thereby.  The arising obstacles are related to the fact that in these cases the Klein-Gordon operator cannot be factorized to the product of two Dirac operators as in the free case (\ref{factor}). Specifically, in the presence of electromagnetic field with 4-potentials  $A^\mu= \{\Phi,\vec A\}$ the {\it squared DE} for the 2-component spinors $a(X), b(X)$ looks as follows (see, e.g., ~\cite[p. 101]{Schweber}, ~\cite[ch.II, sect. 12]{Akhiezer}):
\be{quadrDirac}
(\square_{gen} -m^2 +\vec \sigma (\vec H-i  \vec E) ) a = 0,~~(\square_{gen} -m^2 + \vec \sigma (\vec H+ i \vec E) ) b = 0  
\ee 
and, on account of the last matrix-valued terms, does not allow for interpretation of the $a$ or $b$ components as scalar fields. In (\ref{quadrDirac}) the first term $\square_{gen}$ represents the ordinary Klein-Gordon operator in an external field, and $\vec H, \vec E$ -- the fields themselves, magnetic and electric, respectively. 

Remarkably, an attempt to describe the electron-positron field by the squared DE for only one of the 2-spinors has been undertaken in the paper of Feynman and Gell-Mann~\cite{GellMann}, see also~\cite{Deriglaz}.  In essence, such a description is equivalent to the canonical one (see, e.g., ~\cite{Braun}), since the second 2-spinor can be restored through the procedure of differentiation analogous to that above presented. Nonetheless, such a description turns out to be appropriate in the framework of Feynmann formalism of path integration~\cite{Feynman}. 

On the other hand, the scalar nature of the fields generating the DE solutions reveals itself  when the external (complexified) electromagnetic field is {\it (anti) self-dual}. In fact, the (anti) self-duality conditions 
$\frac{i}{2}\epsilon_{\mu\nu\rho\lambda} F^{\rho\lambda} =\pm F_{\mu\nu}$ in the 3D-form take just the form
\be{selfgual}
\vec H \pm i \vec E =0,~~ (\prt_t \vec A - \nabla \Phi \pm i \nabla \times \vec A =0),  
 \ee 
and their fulfilment guarantees the fulfilment of homogeneous Maxwell equations, for real and imaginary part of the complex strengths separately ~\footnote{{\it Each} solution of Maxwell equations together with its dual is generated by a solution of the complex self-duality constraint, see, e.g., ~\cite{IJGMMP,Sing}}. In the case when, say, the conditions of {\it self-duality} $\vec H - i E =0$ do hold, the squared DE reduces to the ordinary KGE in an external field with complex potentials $A_\mu$ for a doublet of fields $a$, which can then be considered as scalars. These fields after the differentiation  (generalizing (\ref{second2sp})) 
\be{gen_b}
 b=\frac{i}{m} W_{gen} b,
 \ee
define the second pair of functions $b$ which, together with $a$, represents a solution to the DE in an external field. Here $W_{gen}$  is a generalized, in a usual way (that is, by ``prolongation of derivatives''), Weyl operator (\ref{weyl}). Note that since the generalized Weyl operators do not longer commute $\tilde W_{gen} W_{gen} \ne W_{gen} \tilde W_{gen}$,  functions $b$ in (\ref{gen_b}), contrary to $a$,  do not satisfy the KGE in an external field. For {\it (anti) self-dual} external fields functions $a$  and $b$ exchange their places (together with corresponding reflection of spatial coordinates). 

During realization of the above described conception, the canonical problem of relativistic QM  on definition of the states of the electron in external electromagnetic fields (a complete review see, e.g., in ~\cite{Bagrov}) requires essential reformulation. Particularly, in the relativistic  problem of the hydrogen atom, instead of the standard Coulomb  potential, one should use the combined potential $\{\Phi=q/r,~A_\varphi = i \Phi~ tg(\theta/2), ~~ A_r=A_\theta =0\}$. This results in complex self-dual fields whose real part corresponds to electric Coulomb, while imaginary -- to magnetic monopole distributions. Just for such an ansatz the hydrogen atom problem reduces to resolution of a ``pure'' KGE in the joint concordant field of electric and (imaginary) magnetic monopoles~\footnote{In this case the field strengths themselves, contrary to the potentials, do not enter the principal equation (\ref{quadrDirac})}. 

Analogous situation arises in the relativistic prtoblem on the states of particles in a constant and  homogeneous magnetic field. To reduce the DE to the KGE one needs to supplement this field by an {\it imaginary electric} one such that the full complex field be self-dual. Note that this situation can be closely connected with the known problem of imaginary electric dipole moment whose existence inevitably follows from the DE but whose physical status still remains unclear. 

A crucial for the presented scheme question consists, however, in the distinctions of {\it energy spectra} of electrons in complex self-dual fields from the canonical ones, in the usually considered real-valued fields. This problem certainly requires a careful  investigation. 

As for the case of a of gravitational field, the situation here looks much analogous to the electromagnetic case. Specifically, reduction of the DE to the KGE turns out to be possible only in {\it complexified} space-times with complex (anti) self-dual curvature tensor. Such ``(right-) left-flat'' spaces were studied, in particular, in~\cite{Newman} and other papers in connection with the problem of the ``nonlinear graviton'', generalizatioins of twistor theory, etc. Nonetheless, many principal problems arising in this approach, to the first turn those  about the ways of transition to a real physical metric from the initial complex one, are far from being solved. 

\section{Massless case: equivalence of the Weyl and d'Alembert equations}
For completeness, we briefly review here close relations of solutions to the 2-spinor Weyl equation and d'Alembert (wave) equation for a one-component complex scalar field (for detail, see \cite{Sing,Equiv}). 

Consider a massless Weyl equation in the matrix form
\be{WE}
\tilde W \psi =0, 
\ee
where $\tilde W$ is the (conjugate) Weyl operator (see (\ref{weyl})) 
\be{Weyl2}
\tilde W:=(\prt_t +\vec \sigma \nabla)
\ee      
and $\psi$ is a 2-spinor with complex components, say, $\psi^T = \{\alpha, -\beta\}$. 

Making use of the matrix space-time coordinates 
\be{spincoord2}
X=X^+ = \left( \begin{array}{cc}
u & w \\
\bar w & v
\end{array}\right), ~~~
u,v=t\pm z,~~\bar w, w:=x\pm iy,
\ee
one can write out (\ref{WE}) as a simple system of two equations, 
\be{WC}
\begin{array}{cc}
\prt_u \alpha = \prt_{\bar w} \beta, \\
\prt_w \alpha = \prt_v \beta.
\end{array}
\ee

In view of the first equation (\ref{WC}), there exists (locally) a complex function 
$\mu(X)$, for which 
\be{potn1}
\alpha = \prt_{\bar w} \mu, ~~\beta = \prt_u \mu.
\ee
Then the second equation (\ref{WC}) will be identically satisfied if the {\it potential function} $\mu$ is subject to d'Alembert equation
\be{1CKG}
\square \mu:=( -\prt_u\prt_v + \prt_w\prt_{\bar w}) \mu \equiv 0.
\ee
Evidently, a symmetric potential function $\nu(X)$ subject to d'Alembert equation $\square \nu = 0$ can be introduced on the base of the second equation  (\ref{WC}). 

 Thus, {\it any solution to the 2-spinor Weyl equation can be obtained from a solution to one-component d'Alembert equation}. However, in order to represent the principal
expressions in the matrix form, one can use, say, instead of (\ref{potn1}):
\be{potn2}
\alpha=\prt_v \nu - \prt_{\bar w} \mu, ~~~\beta = \prt_w \nu - \prt_u \mu, 
\ee
where now both potentials $\nu,\mu$ should satisfy the d'Alembert equation. Then the last solution to initial Weyl equation can be written in the matrix form:
\be{WS}
\psi = W \zeta, 
\ee
where $W=\prt_t - \vec \sigma \nabla$ is the Weyl operator, and the ``potential row'' is given as $\zeta^T = \{\nu, \mu \}$. Since the latter does always exists (locally), one immegiately checks that the Weyl equations do hold,   
\be{WCH}
\tilde W \psi = \tilde W W \zeta \equiv - \square \zeta =0. 
\ee
      
 Matrix form for potentials is convenient to observe the (restricted) gauge symmetry of the Weyl equation under gradient-wise transformation of potentials of the form  
 \be{WG}
 \zeta \mapsto \zeta +  \tilde W \kappa, 
\ee
 with $\kappa$ being an arbitrary (and independent on $\psi$) column with two components subject to d'Alembert equation. 
 It follows  then from (\ref{WS}) that such a transformation of potentials preserves the initial solution $\psi$ of the Weyl equation. We see thus that {\it free Weyl equation possess a (restricted) gauge symmetry}, in full analogy with the case of massive Dirac equation (see (\ref{grad})).
 
 Another properties of free DE discovered in the paper, among them the existence of chains of solutions (Section 4), of non-canonical scalar-like symmetry transformations (Section 5), of ``positive definite ``energy'' density of Dirac field (Section 6) etc.,  are also in a close analogy with those for massless Weyl equation (see (\cite{Sing}, \cite{Equiv}) for detail). 
 
 Note also that solutions to the latter are in full correspondence with those of electromagnetic {\it (anti-) self-duality constraint} and, via the latter, -- to the solutions of free Maxwell equations (see Section 7). Here we only present a {\it Coulomb-like solution} to Weyl equation (\cite{Sing}),
 \be{WCoul}
 \alpha = \frac{1}{2r}, ~~\beta = - \frac{\mu}{2r}
 \ee 
 which can be obtained in the above described way from the solution to d'Alembert equation 
 \be{dACoul}
 \mu:=\frac{\bar w}{z+r} = \tan{(\frac{\theta}{2})} e^{i\varphi}
 \ee
 corresponding to the {\it stereographic projection} $S^2 \mapsto \mathbb C$.   
Corresponding solution to free Maxwell equations is just the Coulomb one, with a $\delta$-type singularity in origin.

\section{Conclusion}
It has been shown  that free  DE is, essentially, no more than an {\it identical interdependency} between derivatives of a doublet of the Klein-Gordon field. An alternative method to demonstrate the equivalence of Dirac and 2-component Klein-Gordon equations has been elaborated in  ~\cite[Sect. 5A,5B]{Deriglaz}. 
In our treatment, generating solutions to the KGE manifest themselves as potentials for the Dirac ``field strengths''. One reveals thus that free {\it Dirac field is gauge in nature and closely resembles the Maxwell field}.

We have discovered the property of form-invariance of the DE under non-canonical Lorentz transformations in which the generating doublet of Klein-Gordon fields behaves as a pair of {\it scalars} while the second doublet supplementing the first one to a solution of the DE, transforms according to a {\it nonlinear} representation of the Lorentz group. This leads, in particular, to elimination of the generally accepted spinorial 2-valuedness in the 3D rotations. It is known that  the transformation properties of just the {\it free} Dirac/Klein-Gordon fields  predetermine distinct procedures of their secondary quantization in the framework of QED. Therefore, the above presented results force one to put in doubt the competence of the  canonical quantization procedure and ponder over its possible reformulation and/or reinterpretation.   

 In this connection, it is worth noting that A. Zommerfeld ~\cite[ch. 6, sect. 6]{Zommerfeld} had proposed another way to regard the Dirac wave functions as scalars instead of generally accepted (bi)spinor transformations. His proposition makes use of the 4-vector law of transformation of the Dirac $\gamma$-matrices themselves. This procedure preserves defining commutation relations (\ref{commut}) for these matrices and all the principal consequences of the Dirac theory as a whole (see also ~\cite{Zommer}).    
 
As another consequence of the discovered non-canonical links between the fields of Dirac and Klein-Gordon one can distinguish the possibility to obtain a wide class of (singular) solutions to the DE and KGE by subsequent differentiation of (one or two) starting KGE solutions. Physical interpretation of the obtained solutions is generally not evident; however, it would be wrong to ignore the very existence of such entities. Among these, of a special interest is the ``spinor'' analogue (\ref{YukawaSpinor}) of the Yukawa potential and the second static solution to  DE (\ref{SterCoul}) with a string-like singularity. These solutions can perhaps be explicitly interpreted as {\it Dirac fields produced by fermions of two types}. Also remarkable are, certainly, the stationary ``partners'' of these static solutions, especially the solution (\ref{spinorBroglie}) with a finite value of the field charge. 
Corresponding chains of solutions to KGE-DE can be, o course rather speculatively, treated as {\it excited states} (``resonances''). Probably, it will be not hard to form a complete list of stationary singular solutions to KGE-DE.   

One can obtain an interesting variety of these solutions  by a {\it complex shift}  $z \mapsto z + i a$ along the coordinate corresponding to the symmetry axis $(Z)$ of the initial solution. The so transformed solutions, say, (\ref{Yukawa}) or (\ref{BroglieKG}) acquire then a {\it ring-like} singularity of a radius $a$, as in the case of above-mentioned Kerr-Newman solution in GTR~\cite{Witt}. 
It is well known that the Kerr-Newman solution has some properties analogous to those of Dirac particle (in particular, the gyromagnetic ratio for this solution is $g=2$~\cite{Carter}). Therefore such a deformed DE solution could be very interesting, say, in the context of the {\it Dirac -- Kerr-Newman electron} model developed by A.Ya. Burinskii~\cite{Burin}. 

Finally, it is worth to note that the considered scheme allows to bypass the well-known Pauli  theorem ~\cite{Pauli} on the sign-indefinite energy density for the fields of half-integer, and charge density -- of integer spins. Actually, as we have seen, any DE solution can be equipped with a positive definite energy density, and any KGE solution -- with a  positive definite charge density -- probability density~\footnote{Thus, the known motivation of Paul Dirac~\cite{DiracMotiv} in his search of a novel relativistic equation with positive definite probablility density turns out to be in a sense superfluous!}.

On a whole, however, we did not claim here to suggest some new physical theory, interpretation or quantization procedure. We desired only to simply and rigorously  demonstrate that a number of paradigmatic settings dominating at present within the (firstly quantized) relativistic field theory are in fact controversial  (questionable) and actually should be reconsidered. For this  it is certainly necessary to elaborate a number of supplementary investigations, especially for the case when external fields, electromagnetic and gravitational, are present.

\section*{Acknowledgments}
The authors are grateful to Igor V. Volovich,  Victor V. Zharinov and participants of scientific seminars in the Steklov Mathematical Institute they supervise for valuable comments.

 \end{document}